\documentstyle[preprint,aps,epsf]{revtex}                  
\global\let\epsfloaded=Y 
\begin{document}
\pagestyle{empty}                                      
\preprint{
\font\fortssbx=cmssbx10 scaled \magstep2
\hbox to \hsize{
\hbox{
            }
\hfill $
\vtop{              
 \hbox{ }}$
}
}
\draft

\begin{center}
{\large {\bf Erratum and corrigendum:\\
Interactions of neutrino with 
an extremely light scalar}\\ (Phys. Lett. {\bf B444}, 75(1998))}
\\
{Xiao-Gang~He\footnote{email:hexg@penguin.phys.ntu.edu.tw}$^{1,2}$, 
Bruce H J McKellar\footnote{email:mckellar@physics.unimelb.edu.au}$^{1,3}$ 
and G J Stephenson Jr\footnote{email:gjs@baryon.phys.unm.edu}$^{1,4}$}
\\
{
$^1$ School of Physics, University of Melbourne, Parkville, Vic. 
Australia 3052\footnote{X-G. H. and G. J. S. Jr. visited Melbourne 
while the original paper was constructed}\\
$^{2}$Department of Physics, National Taiwan University,
Taipei, Taiwan 10717\\
$^3$ On leave at Special Research Center for the Subatomic Structure of 
Matter\\
University of Adelaide, SA,  Australia 5005 
\footnote{at the time of the original paper}\\
$^4$ Department of Physics and Astronomy\\
University of New Mexico, Albuquerque, NM USA 87131}
\end{center}

%
%
\vspace{1cm}

Chang, Keung and Yeh [\ref{Chang:2003ww}] have proved that Goldstone scalars of the type we used in our paper can couple to neutrinos only via $\gamma_5$ coupling, contrary to the result we obtained.

Our incorrect result was a consequence of some errors  in sections III and IV. Here we show how to correct them and satisfy the general theorem of ref [1].  We then show how to extend our model to generate the required coupling, ensuring that the coupling strength and the scalar mass are both small as desired.

In section III, eq. (12) should be: $\phi = N[2 v_\chi^2 v_H Im(h^0) 
+ v^2_H v_\chi Im(\chi) + (v^2_H + 4 v_\chi^2) v_s Im(S)]$, and 
the $2-2$ entry in the coupling matrix of eq. (14) should be: 
$+M (v_H^2+4 v_\chi^2)$.
With these corrections, the scalar coupling of $\phi$ to $\bar \psi \psi$ 
is exactly zero which agrees with the conclusion in Ref.[1]. 

In section IV the new global $U(1)$ charges in eq. (19) 
for $l^i_L$ and $\nu_R^i$ should be
$\alpha_i$ and $-\alpha_i$, respectively.
The $2-2$ and $3-3$ entries in eq. (21)
should be changed to $-m_{22}v_{H_1}^2$ and $-m_{33}v^2_{H_1}$.
Then, as in  section III, the model
does not have a non-zero scalar $\phi$ to $\bar \psi \psi$ coupling as it is.

It is however still feasible to generate the required coupling in a majoron model, and to ensure that it is small.  After all we still have to generate a non-zero mass for the majoron.  We pointed out in the paper that 
it is possible to use dimension 5 operators, which may be induced by gravity, to generate the majoron mass, and give such a term in eq. (17).  Such a coupling can also induce  a non-zero 
scalar coupling, but this scalar coupling is too small to generate a $g$ 
in the range of $10^{-15} \sim 3\times 10^{-21}$. However, a similar gravity  induced explicit lepton number violating operator can exist and may be used to generate the desired scalar couplings. For example, a coupling of the form
$(\sigma/m_{pl})\bar \nu^C_R \nu_R S^\dagger S + H.C.$, 
which was proposed by  Akkhmedov,  Berezhani and Senjanovic [\ref{Akhmedov:1992hh}], will induce
a non-zero $g$ with $g\sim (\sigma /m_{pl})(v_s/M) (m^2_D/M)$. For 
$m_\nu \sim m^2_D/M \sim 1$ eV, and $v_s/M \sim 10^{-6}$, it is possible to 
obtain the desired scalar coupling. 

In the Dirac neutrino model,
a non-zero scalar coupling can be generated by introducing global $U(1)$ 
breaking operators. For example a term of the form 
$(\sigma/m_{pl}) \bar l_{L1}\nu_{R_1} 
H_1 S + H.C.$ can generate a $g$ with $g \sim (\sigma /m_{pl}) 
(v_1^3/(v^2_1+v^2_2))$ which can be made to be in the desired range.

We thank D. Chang and W.-Y. Keung for helpful discussions.

\noindent
\section*{References}

\begin{enumerate}
\item \label{Chang:2003ww}
D.~Chang, W.~Y.~Keung and C.~P.~Yeh,
Phys.\ Rev.\ D {\bf 67}, 055007 (2003).
[arXiv:hep-ph/0211016]. 

\item \label{Akhmedov:1992hh}
E.~K.~Akhmedov, Z.~G.~Berezhiani and G.~Senjanovic,
Phys.\ Rev.\ Lett.\  {\bf 69}, 3013 (1992)
[arXiv:hep-ph/9205230].

\end{enumerate}

\newpage

\vfill
\title{
Interactions of a neutrino with an extremely light scalar 
}

\vfill
\author{
Xiao-Gang He$^{1,2}$, Bruce H J McKellar$^{1,3}$ and 
G J Stephenson Jr $^{1,4}$
}
\address{
\rm $^1$School of Physics, 
University of Melbourne, 
Parkville, Vic. AUSTRALIA 3052 
\\
\rm
$^2$Department of Physics, National Taiwan University,
Taipei, Taiwan 10717
\\
\rm $^3$On leave at Special 
Research Centre for the Subatomic Structure of 
Matter\\ University of Adelaide, 
SA.  AUSTRALIA 5005
\\
and
\\
$^4$Department of Physics and Astronomy\\
University of New Mexico, 
Albuquerque, NM USA 87131 
\vspace{0.5cm} 
}

%
%
\vfill
\maketitle
\begin{abstract} 
An extremely light scalar weakly interacting with light 
neutrinos has   
interesting consequences in stellar evolution, neutrino 
oscillations and  
laboratory neutrino mass measurements. In this paper we 
construct realistic  
gauge models for such scalar-neutrino interactions.   
\end{abstract} 
 
\newpage 
\section{Introduction} 
Recently, it has been shown that, if neutrinos weakly interact
with an extremely light scalar boson, it is possible to introduce
additional MSW like effects in the sun which alter the details of
{\it r}-process nucleosynthesis in supernovae, to impact the
earliest formation of stars, and to modify data near the endpoint
of the beta spectrum in laboratory measurements of neutrino mass
which may aid in the resolution of the ``negative mass-squared
problem''~\cite{gms}

The specific renormalizable Lagrangian introduced whose 
phenomenological consequences are discussed in ref\cite{gms} 
is of the form 
\begin{eqnarray} 
L = \bar \psi (i\partial - m_\nu) \psi + {1\over 2} [\phi  
(\partial^2 - m_s^2)\phi] 
+g\bar \psi \psi \phi\; \label{lag} 
\end{eqnarray} 
where $\psi$ and $\phi$ are the neutrino and scalar boson, 
respectively. 
In order not to be in conflict with other constraints, the 
scalar mass 
has to be around\cite{gms} 
 $1.3 \times 10^{-18}$ eV, and the coupling $g$ is in the 
range $3\times 10^{-21} < |g| < 10^{-15}$.  
 
In general the interaction between neutrinos and scalar can also 
have  
a renormalizable term of the form 
\begin{eqnarray} 
L = g' i\bar \psi  \gamma_5 \psi \phi\;.\label{g5} 
\end{eqnarray} 
This type of interaction, however, will not affect the results 
obtained in 
ref.\cite{gms} because, in the macroscopic neutrino background 
considered there, the 
average of spin 
of the neutrinos is zero and the interaction term in Eqn 
(\ref{g5}) will have 
no effect. 
 
In this paper we study models which extend the standard model 
by the introduction of a neutrino 
interaction with an 
extremely light scalar. In section 2, we will discuss some basic 
features  
and difficulties  
of such interactions in gauge models based on the Standard Model 
(SM). 
A toy model will be given  
 which also serves to introduce our notation. In section 3, we 
will discuss the possibility of realizing such interactions in 
Majoron models, 
and in section 4, we will discuss the possibilities of realizing 
such 
an interaction in models with only Dirac neutrinos. 
 
\section{A simple model} 
 
In the minimal SM it is not possible to accommodate 
the interactions of Eqn (\ref{lag}) although there is a scalar, 
 the Higgs boson, in 
the model because this scalar does not couple to 
neutrinos and,  
further, experimental data constrain its mass to be larger than 
$77.5$ GeV\cite{higgs}. 
 
In order to get a new scalar interaction like that of Eqn 
(\ref{lag}), one must go beyond the standard model, while still 
respecting 
the $SU(2)_L\times U(1)_Y$ transformation properties of the 
existing particles of the standard model.  These transformation 
properties of leptons and the Higgs boson, 
which are relevant to our discussion, under  
the 
SM gauge group $SU(2)_L\times U(1)_Y$ for electroweak 
interactions are given  
by, 
 
\begin{equation} 
L_L^i = \left ( \begin{array}{l} \nu\\ e \end{array} \right 
)_L\;:  
(2,-{1\over 2})\;,\;\; 
E_R^i = e^i_R\;:(1,-1)\;\;\; , \mbox{and} \quad
H = \left ( \begin{array}{l} h^0\\h^- \end{array}  
\right ) \;:(2,-{1\over 2})\;, 
\end{equation} 
where $i = e, \mu, \tau$ is the generation index. 
  
\relax From the quantum numbers of the particles, it is clear  that  
one must introduce 
new particles to generate the interactions we want.  The 
extremely 
light  
scalar which weakly couples to neutrinos 
will be a new particle beyond the minimal SM. 
The simplest way to do this is 
to introduce a real scalar $S$  
which transforms under the SM gauge group as a singlet $(1,0)$.  
This scalar will not couple to the left-handed neutrinos  
by any renormalizable Lagrangian, so we also  
 introduce right 
handed neutrinos $\nu_R^i$ which transform as singlets 
$(1,0)$, and have a Majorana mass matrix which is introduced as 
a free parameter.  
With these new particles, it is possible to have interactions 
between   
the new scalar  and the neutrinos. Including the Higgs 
interactions, the relevant terms are 
 
\begin{eqnarray} 
L_{new} &=& -\bar L_L {m_\nu^\dagger \over v_H} H^\dagger \nu_R 
-  
\bar \nu_R {m_{\nu }\over v_H} H L_L 
- {1\over 2} \bar \nu_R^c M^\dagger \nu _R - {1\over 2} \bar 
\nu_R M\nu_R^c 
\nonumber\\ 
&&+ \bar \nu_R^c C^\dagger \nu_R S + \bar \nu_R C\nu_R^c S 
+ {1\over 2} \partial_\mu S \partial^\mu S + V(S)\;, 
\label{simpmod1}  
\end{eqnarray} 
where $v_H$ is the vacuum expectation value (VEV) of $H$, and 
$V(S)$  the scalar field potential which has the form 
\begin{eqnarray} 
V(S) = -{1\over 2} m_s^2 S^2 + \mbox{ other terms up to fourth 
power in S}\;. \label{simpmod2} 
\end{eqnarray} 
The generation index on leptons has been suppressed.  
This Lagrangian of Eqn.(\ref{simpmod1}, \ref{simpmod2}) 
contains  some free  parameters:  
a scalar mass $m_s$, a scalar-neutrino coupling matrix $C$,  
a Majorana mass matrix  $M$ for $\nu_R$, and a Dirac mass matrix  
$m_\nu$.  
Notice that lepton number is explicitly broken in this model. 
In the neutrino mass eigenstate  
bases, one then has 
\begin{eqnarray} 
L &=& {1\over 2} \bar \psi i\partial \psi -{1\over 2} \bar \psi 
\hat M \psi 
+ \bar \psi_L \hat C \psi_R S + \bar \psi_R \hat C^\dagger 
\psi_L S 
+ {1\over 2} (\partial_\mu S \partial^\mu S - m_s^2 s^2) + 
...\;, 
\end{eqnarray} 
where  
$\hat M$ is the diagonal mass matrix, $U$ is 
a unitary matrix which diagonalizes the mass matrix and 
transforms the coupling matrix $C$ to a new matrix 
$\hat C$ in the 
following way 
 
\begin{eqnarray} 
\hat M &=& U^T \left (  
\begin{array}{ll}  0& m_\nu^{T}\\ m_{\nu}& M 
\end{array} \right ) U \;,\;\; 
\hat C = U^T \left ( \begin{array}{ll} 0&0\\0&C \end{array} 
\right ) U\;. 
\end{eqnarray} 
The mass eigenstate is  
$\psi= \psi_L +\psi_R$, with 
$\psi_L = N_L$, $\psi_R = N_L^c$, where 
 
\begin{eqnarray} 
N_L &=& \left ( \begin{array}{l}  N_{1L}\\ N_{2L}  \end{array} 
\right ) 
= U^\dagger \left ( \begin{array}{l} \nu_L \\ \nu_R^c 
\end{array} \right )\;, 
\end{eqnarray} 
 
In this simple model the masses are all free parameters. Using 
this freedom, 
it is  
possible to obtain the parameter ranges used in ref~\cite{gms}.  
To see how this happens, 
consider a  
simple case with just one generation. We assume that the light  
neutrino mass is due to the See-Saw mechanism\cite{see-saw}, 
so that 
$M \gg m_\nu$. The two mass eigenvalues are: $m_l \approx 
-m_{\nu}^{ 2}/M$ and  
$m_h \approx M$, 
and $U$ is given by 
\begin{eqnarray} 
U = \left ( \begin{array}{ll} \cos\theta& -\sin\theta\\ 
\sin\theta& \cos\theta \end{array} \right )\;,\;\; \sin\theta 
\approx -{m_\nu  
\over M}\;. 
\end{eqnarray} 
The scalar coupling to the light neutrino is given by 
\begin{eqnarray} 
g = \sin^2\theta C \approx {\left (m_\nu \over M \right )^2} C 
\approx - {m_l\over M} C\;. 
\end{eqnarray} 
The upper limit of the 
electron neutrino mass is constrained to be of order ten 
eV\cite{pdg}\footnote{ 
In the presence of a neutrino cloud, the vacuum neutrino mass 
can be  
larger than the experimental bound, which bounds the effective 
mass inside the cloud\cite{gms}.}. A value of  
$g$ in the range of $ 10^{-21} \sim 10^{-15}$, follows from 
values of   
$M$  in the range of $10^{12} \sim 10^{6}$ GeV if $C$ is of 
order one. 
 
This exercise is just a demonstration that it is indeed possible 
to have  
a model which satisfies the requirements of ref\cite{gms}. 
The coupling  
between the scalar and the  
light neutrino is naturally small. However, although it 
is possible to have a small scalar mass due to the arbitrariness 
of the parameters, 
it will be  
 much more interesting if some mechanism can be found which 
guarantees  
the lightness of the scalar particle.  
 
If the scalar is a Goldstone boson resulting from the breaking 
of a  
global symmetry, the 
scalar is naturally light, $m_s = 0$.  
A small mass can develop for the Goldstone boson 
if the symmetry is explicitly broken  
weakly due to  
some mechanism.
The Axion\cite{axion}  
is a famous example of this mechanism. However, available 
Axion  
models have  masses for the spin zero particle which are too 
large. In the 
following we  
consider models in which possible global symmetries are broken  
spontaneously as well  
as explicitly due to gravitational effects, such that the 
would-be  
Goldstone boson resulting from spontaneous symmetry breaking 
receives a 
small mass. 
 
\section{\bf A Majoron model} 
 
In the example given before, the lepton number is explicitly 
broken by Majorana 
mass terms and the scalar coupling terms. We now consider a 
model in 
which  
lepton number is spontaneously broken with a massless Majoron. 
The model to be studied contains in  
addition to the usual minimal SM particle contents, right-handed 
singlet neutrinos $\nu_R^i$,  
one triplet Higgs $\chi$ and a  
singlet scalar $S$. The transformation properties under the 
$SU(2)_L\times U(1)_Y\times 
U(1)_{lepton}$ group  for scalar particles are: 
 
\begin{eqnarray} 
H: (2,-1/2)(0)\;;\;\;\chi : (3,1) (-2)\;;\;\; S: (1,0)(-2)\;. 
\end{eqnarray} 
Here the quantum number in the last bracket is the lepton 
number. Lepton number 
is spontaneously broken when $\chi$ and $S$ develop non-zero 
vacuum expectation values  
$v_\chi$ and $v_{S}$, respectively.  
The Majoron field is given by 
\begin{eqnarray} 
\phi = N [ 2 v_\chi^2 v_H \Im(h^0) - v_H^2 v_\chi \Im(\chi)  
-(v_H^2 + 4 v_\chi^2)v_S  
\Im S]\;, 
\end{eqnarray} 
where $N =1/ 
\sqrt{(2v_\chi^2v_H)^2+(v_H^2v_\chi)^2+((v_H^2 + 
4v_\chi^2)v_S)^2}$  
is the normalization constant.  
 
The mass matrix for the neutrinos in  
the bases $(\nu_L, \nu^c_R)$ is given by, 
\begin{eqnarray} 
M_{neutrino} = \left ( \begin{array}{ll} 
m&m_{D}^{ T}\\ 
m_D& M \end{array} \right )\;, 
\end{eqnarray} 
where $m \sim v_\chi$, $m_D\sim v_H$ and $M\sim v_S$ are the 
Majorana mass matrix for the left-handed  
neutrinos, the Dirac mass matrix for neutrinos and the Majorana 
mass matrix  
for the right-handed neutrinos, respectively.  
 
The Majoron- neutrino coupling matrix is given by 
\begin{eqnarray} 
C = {N_\psi\over \sqrt{2}}\left ( \begin{array}{ll} 
-m v^2_H & 2m_D^{T}v_\chi^2\\ 
2m_Dv^2_{\chi}& - M(v_H^2 + 4 v_\chi^2) 
\end{array} 
\right ) 
\end{eqnarray} 
For simplicity we work with just one generation and assume $m$, 
$m_D$ are real 
and $M$ has a phase of\footnote{Without complex phases, the 
Majoron 
being a pseudoscalar couples to neutrinos through the $\gamma_5$ 
coupling of 
Eqn. (\ref{g5}).  
Complex phases in the interaction are needed to induce the 
scalar coupling of  
Eqn. (\ref{lag}). We choose a phase of $\pi/2$ to maximize the 
effect.} 
$\pi/2$.  
It has long been known that a Majoron  
model of the type discussed here  
without the singlet is ruled out by experimental data from LEP 
on Z decay  
width\cite{pdg},  
because of additional light charged scalar decay modes for 
Z\cite{majoron}. This  
problem can be solved by the introduction of the singlet scalar, 
with  
$v_S \gg v_H \gg v_\chi$ such that the light charged scalar 
decay 
modes of Z 
are suppressed. 
We will assume this is the case in the discussion to follow.  
 
After diagonalization of the neutrino mass matrix and in the 
mass 
eigenstate 
bases, the light neutrino mass and interaction with the scalar 
is given by 
\begin{eqnarray} 
L = -{1\over 2} \bar \psi m_l \psi - {1\over 2\sqrt{2}v_S} 
[i \bar \psi (m_1 + {2m_D^2 \over M} \sin \delta_1)\gamma_5 \psi 
-{2m_D^2 \over M} \cos\delta_1 \bar \psi \psi]\phi\;, 
\end{eqnarray} 
where $m_1 \approx \sqrt{m^2+(m_D^{2}/M)^2}$ is the eigenmass 
and  
$\delta_1 = \mbox{arctg}(m_{D}^{2}/mM)$.  
 
As has been discussed before, the interaction term with 
$\gamma_5$ does 
not affect the results of interest to us, so we can ignore  it 
in the remainder of this section. We 
then identify 
\begin{eqnarray} 
g = {m_D^2 \over \sqrt{2}Mv_S} \cos\delta_1\;. 
\end{eqnarray} 
 
It is easy to obtain a neutrino  mass 
close to the upper limit of order ten eV. Also 
if we assume $m$ is the same order of magnitude as $m_{D}^{ 
2}/M$, 
$\cos \delta_1$ is of order one, $M\sim v_S$ is in the 
range of $10^6$ to 
$10^{12}$ GeV, and the coupling $g$ is in the range of  $10^{-15} 
\sim 10^{-21}$.

So far the Majoron mass is exactly 
zero. 
However, this may not hold when gravitational effects 
are
considered.
It has been argued that, in the presence of 
gravity,
the lepton number symmetry may be explicitly broken and 
result
in a small
mass for Majoron\cite{abms}. The mass would be 
inversely
proportional to
the Plank mass and therefore can be 
made very small.
The lowest terms have dimension five. To this 
order, a
number of terms can
contribute. For example a term 
of the following 
form can appear 
in the 
potential

\begin{eqnarray}
V_{add} = \beta {1\over 
m_{Pl}}(\chi^+\chi)^2 S + H.C.\;,
\end{eqnarray}
where $m_{Pl}$ 
is the Planck mass. This term satisfies the
condition that 
when 
the Planck mass goes to infinity, its contribution
vanishes.
Due 
to the appearence of this term, the minimal conditions for
the 
potential
change, and the Majoron $\phi$ will mix with the 
massive scalar
and also become 
massive. The mass-squared 
$m_\phi^2$ is given by

\begin{eqnarray}
m_\phi^2 = -\beta 
{v^4_\chi \over 2\sqrt{2} v_s m_{Pl}}\;.
\end{eqnarray}

This 
mass can be easily made to be around $10^{-18}$ eV with
$\beta$ 
to be
of order one.
The analysis with
additional terms can be 
easily carried out in the same way.

\section{\bf A model for 
Dirac neutrino -- scalar interactions}

The conditions required 
for neutrino-scalar coupling of the type
of Eqn (\ref{lag}) can 
also be
realized 
for Dirac neutrinos. In the following we 
present a simple 
model to demonstrate this fact. In this 
model
we introduce a new global $U(1)$ 
symmetry whose quantum 
numbers for different fields 
are indicated in the second 
bracket in the following,

\begin{eqnarray}
l^i_L&:& 
(2,-1/2)(0)\;;\;\;\nu_R^i : 
(1,0)
(\alpha_i)\;,\nonumber\\
H_j&:& (2,-{1\over 
2})(\beta_j)\;;\;\;S: (1,0)(1)\;; \nonumber\\
\alpha_1 &=& 
0\;;\;\; \alpha_2 = \alpha_3 = 1\;;\;\; \beta_1 
=
0\;;\;\;
\beta_2 = 1\;.
\end{eqnarray}
Here $i$ is the lepton 
generation index and $j$ is doublet Higgs
index, which is 
introduced because we find that two Higgs
doublets are 
needed.
The global symmetry is broken by VEV's of $H_j$ and $S$. 
The
corresponding 
Goldstone boson
is given 
by
\begin{eqnarray}
\phi = N [v_{H_2}^2 v_{H_1} \Im(h_1^0) - 
v_{H_1}^2 v_{H_2} \Im
(h^0_2) -
(v_{H_1}^2+v_{H_2}^2) v_S 
\Im(S)]\;,
\end{eqnarray}
where $N = 
1/\sqrt{(v_{H_2}^2v_{H_1})^2 + (v_{H_1}^2v_{H_2})^2 
+
((v_{H_1}^2+v_{H_2}^2)v_S)^2}$ is the normalization 
constant.

The mass matrix and $\phi$ neutrino coupling matrix 
are given by
\begin{eqnarray}
M &=& \left 
(\begin{array}{lll}
m_{11}&m_{12}&m_{13}\\
m_{21}&m_{22}&m_{23}\\
m_{31}&m_{32}&m_{33}
\end{array}
\right 
) \;,\nonumber\\
C &=& N\left ( 
\begin{array}{lll}
m_{11}v_{H_2}^2&-m_{12}v_{H_1}^2&-m_{13}v_{H_1}^2\\
m_{21}v_{H_2}^2& 
m_{22}v_{H_2}^2&-m_{23}v_{H_1}^2\\
m_{31}v_{H_2}^2&-m_{32}v_{H_1}^2&m_{33}v_{H_2}^2
\end{array}
\right 
)
\end{eqnarray}

The mass matrix can be diagonalized by 
bi-unitary diagonalization, that is
\begin{equation}
\hat m = 
\left 
(\begin{array}{lll}
m_1&0&0\\
0&m_2&0\\
0&0&m_3
\end{array}
\right 
) = V_L M V_R\;,\end{equation}
which gives a transformed, 
coupling matrix
\begin{equation}
\hat C = V_L C 
V_R\;.
\end{equation}

To illustrate this model, consider an 
example with two
generations. The coupling
matrix
is of the 
form
\begin{eqnarray}
\hat C &=& -Nm_{12}(v^2_{H_1} + 
v^2_{H_2})
e^{i\delta_R} \left (\begin{array}{ll}
c_L 
s_R&c_Lc_R\\
s_Ls_R&s_Lc_R
\end{array}
\right )\nonumber\\
V_L 
&=& \left ( \begin{array}{ll}
c_L&-s_Le^{i\delta_L}\\
s_L&c_L 
e^{i\delta_L}
\end{array}
\right )\;,\;\; V_R = \left 
(\begin{array}{ll}
c_R&-s_R\\
s_Re^{i\delta_R}&c_Re^{i\delta_R} 
\end{array}
\right )
\end{eqnarray}
where $c_{L,R} = 
\cos\theta_{L,R}$ and $s_{L,R} =
\sin\theta_{L,R}$.
This model 
is very different from the Majoron model discussed 
previously. 
At least two generations with mixing are needed. The 
Goldstone 
boson coupling to the lightest neutrino is given 
by
\begin{eqnarray}
g = {m_{12}\over v_S} \sin\delta_R 
c_Ls_R\;.
\end{eqnarray}
With $m_{12}$ in the eV range, $c_Ls_R$ 
of order one 
and $v_S$ in the range $10^6 \sim 10^{12}$
GeV, it 
is easy to obtain $g$ in the desired range of $10^{-15}
\sim 
10^{-21}$.

Again the Goldstone boson mass becomes non-zero 
when
gravitational effects
are considered. As an example we 
introduce a term of the form

\begin{eqnarray} 
V_{add} = \beta {(H_2^\dagger H_2)^2 S\over m_{Pl}} + H.C.  
\end{eqnarray} 
in the potential. The field $\phi$ develops a mass given by 
\begin{eqnarray} 
m_\phi^2 = \beta {v^4_{H_2}\over v_s m_{Pl}}\;, 
\end{eqnarray} 
which can easily be in the desired range. 
 
\section{Conclusion} 

Neutrinos that interact weakly with an extremely light scalar have
many interesting consequences in steller evolution, neutrino
oscillations and laboratory neutrino mass measurements. In this paper
we have study the possibility of realizing such interactions in
gauge models.  We have shown that it is indeed possible to
construct such models, and that it is possible to arrange for the
new scalar to be very light and for its coupling to neutrinos to
be naturally very small.  Models which exhibit this behaviour are
constructed in which the neutrinos are either Majorana or Dirac
particles.

\noindent
{\bf Acknowledgement}

This work was supported in part by the Australian Research Council
and by National Science Council of ROC (NSC 87-2811-M-002-046).
BHJM thanks the Department of Physics of the National Taiwan
University, and the National Center for Theoretical Sciences,
Taiwan for hospitality where part of this work was done.  GJS
thanks the Divison of International Programs of
the National Science Foundation for travel
support and the University of Melbourne for its
hospitality.  We thank Ray Volkas for helpful discussions.

\end{document}